\documentclass[sigconf]{acmart}
 
\AtBeginDocument{%
  \providecommand\BibTeX{{%
    \normalfont B\kern-0.5em{\scshape i\kern-0.25em b}\kern-0.8em\TeX}}}

\usepackage{algorithmic}
\usepackage{graphicx}
\usepackage{textcomp}
\usepackage{xcolor}
\usepackage{url}
\usepackage[inline]{enumitem}

\usepackage[]{xcolor}
\newcommand{\TODO}[1]{\textcolor{red}{#1}\GenericWarning{}{LaTeX Warning: TODO: #1}}\newcommand\todo\TODO

\usepackage[flushleft]{threeparttable}
\usepackage{framed}
\setcopyright{acmcopyright}
\copyrightyear{2020}
\acmYear{2020}
\setcopyright{acmcopyright}\acmConference[ICSEW'20]{IEEE/ACM 42nd International Conference on Software Engineering Workshops }{May 23--29, 2020}{Seoul, Republic of Korea}
\acmBooktitle{IEEE/ACM 42nd International Conference on Software Engineering Workshops (ICSEW'20), May 23--29, 2020, Seoul, Republic of Korea}
\acmPrice{15.00}
\acmDOI{10.1145/3387940.3391488}
\acmISBN{978-1-4503-7963-2/20/05}

\begin{document}

\title{On the Relevance of Cross-project Learning with Nearest Neighbours for Commit Message Generation}

\author{Khashayar Etemadi}
\email{khaes@kth.se}
\affiliation{%
  \institution{KTH Royal Institute of Technology}
  \city{Stockholm}
  \country{Sweden}
}

\author{Martin Monperrus}
\email{martin.monperrus@csc.kth.se}
\affiliation{%
  \institution{KTH Royal Institute of Technology}
  \city{Stockholm}
  \country{Sweden}
}

\begin{abstract}
Commit messages play an important role in software maintenance and evolution. Nonetheless, developers often do not produce high-quality messages. A number of commit message generation methods have been proposed in recent years to address this problem. Some of these methods are based on neural machine translation (NMT) techniques. Studies show that the nearest neighbor algorithm (NNGen) outperforms existing NMT-based methods, although NNGen is simpler and faster than NMT. In this paper, we show that NNGen does not take advantage of cross-project learning in the majority of the cases. We also show that there is an even simpler and faster variation of the existing NNGen method which outperforms it in terms of the BLEU\_4 score without using cross-project learning.
\end{abstract}



\maketitle

\section{Introduction}


\label{sec:intro}
Software developers usually use Version Control Systems (VCS) during the development process of software to collaborate and keep track of the changes in the project. In VCSs, each commit contains a set of changes (i.e. a \textit{diff}) and a message (i.e. a \textit{commit message}). The message is written by developers and meant to describe and explain the change. Having high quality commit messages is important for understanding program evolution \cite{cortes2014automatically}, yet many of the messages written by developers are of low quality\cite{liu2019atom}.

In recent years, a number of automatic commit message generation techniques have been proposed to reduce the time needed for producing high-quality commit messages. One of the main research trends in this regard is using Neural Machine Translation (NMT) to translate code diffs to commit messages. In one of the first works on this topic, Jiang et al. \cite{jiang2017automatically} proposed using a RNN Encoder-Decoder architecture for the translation. Their evaluations show that $23.8\%$ of the messages generated using this technique (called \textit{NMT1}) are identical to the actual messages written by developers (also called \textit{reference messages}). Recently, there have been more advanced versions of NMT-based commit message generation methods \cite{liu2019generating, liu2019atom, xu2019commit} that try to improve NMT1, for instance by taking into account the structural information extracted from the code changes.

By investigating the reasons behind the good performance of NMT1, Liu et al. \cite{liu2018neural} conclude that the code diffs of the most high-quality commit messages generated by NMT1 are similar to one or more code diffs in the training set. Based on this observation, they introduce a nearest-neighbor-based recommender technique (NNGen) for commit messages. For a given test diff, NNGen finds the nearest diff in the training set and outputs the reference message of that nearest diff as the generated message for the given diff. Their experiments on the same dataset as the one used by Jiang et al. \cite{jiang2017automatically} indicate that NNGen outperforms NMT1. A more recent study shows that NNGen outperforms other advanced versions of NMT-based commit message generation methods as well \cite{liu2019atom}. 

For evaluations, NNGen \cite{liu2018neural} uses the data collected from top $1k$ Github repositories to generate a commit message for a given diff. However, it is not examined whether there is actually some effective cross-project learning happening. In this paper, we devise and carry out original experiments to investigate this question. As we will see, we report on a negative result: there is no real cross-project learning that improves the results. Our experiments are as follows. First, we detect the project containing the nearest commit for each sample in the testing dataset of NNGen. Second, we introduce and measure the effectiveness of Simple-NNGen, a new variation of NNGen with the difference that it only searches in the same project to find the nearest neighbor for a given diff.

Our experiments on NNGen show that in $60\%$ of the cases the nearest diff is selected from the same project as the test diff project. Interestingly, we also find that when the nearest diffs are selected from the same project, the generated messages have significantly higher quality than when they are selected from other projects.
Second, Simple-NNGen outperforms NNGen in terms of BLEU\_4 score which is a textual similarity metric widely used for assessing commit message generation techniques. Since Simple-NNGen outperforms NNGen and NNGen outperforms NMT-based techniques, this shows that no technique is able to do significant and fruitful cross-project learning. In other words,  although the training dataset used in \cite{liu2018neural} contains commits from $1k$ Github repositories, NNGen performs better when it ignores training commits from $999$ project and selects the nearest neighbor just from the test diff project.  Our results clearly show the relative absence of cross-project learning for commit message generation and call for more research in that field.

To sum up, our contributions are:
\begin{itemize}
\item an original experiment showing that there is little cross-project learning happening for commit message generation based on nearest-neighbors that improves the results.
\item a dataset for commit message generation enriched with additional metadata, this dataset is made publicly available for future research.
\item Simple-NNGen, a simple commit message generation system which beats the state-of-the-art and can now act as baseline in future research in this field.
\end{itemize}

\section{Experimental Methodology}

\subsection{Background on Nearest-neighbour for Commit Message Generation}

\label{sec:nngen}
NNGen, introduced in \cite{liu2018neural}, works as follows: for a given commit diff (i.e. \textit{test diff}), it finds the nearest diff in the training set (i.e. \textit{training diff}) and returns the message of it. NNGen finds the nearest diff with a two-step algorithm as follows:
\begin{enumerate*}
    \item It selects the top five diffs from the training set with highest cosine similarity.
    \item It computes the BLEU\_4 score between the test diff and each of the five selected diffs. The training diff with the highest BLEU\_4 score from the initial five is returned as the nearest diff.
\end{enumerate*}

The automatic evaluation of commit message generation techniques is usually performed by computing the BLEU\_4 score between generated messages and human-written messages \cite{liu2018neural, liu2019atom, jiang2017automatically, liu2019generating, xu2019commit, fraternali1999tools, wei2019code}. The BLEU\_4 score is the product of geometric average of the modified n-gram precisions (represented by $p_i$ in \autoref{eq:bleu}) and the brevity penalty which is a penalty for short messages.

\begin{equation}
\label{eq:bleu}
    BLEU\_4 = BP \times \exp \left(\sum\limits_{1 \le i \le 4} \dfrac {\ln p_i}{4}\right)
\end{equation}

The brevity penalty (BP) is used to avoid the bias in favor of short generated messages. BP is computed according to \autoref{eq:bp} in which $r$ and $c$ stand for the length of reference messages and generated messages, respectively.

\begin{equation}
\label{eq:bp}
    BP=\left\{ \begin{array}{c}
    1\ \ \ \ \ \ \ \ \ \ \ \ \ \ \ c > r \\ 
    e^{1-r/c}\ \ \ \ \ \ c \leq r \end{array}
    \right.
\end{equation}

The BLEU\_4 score is a metric widely used to assess the accuracy of translation techniques \cite{papineni2002bleu}. A higher BLEU\_4 score for a generated message shows that the message is more similar to the one written by the human developer. Liu et al. \cite{liu2018neural} also consider this metric as a textual similarity distance metric used in the second step of NNGen described above.

\subsection{Research Questions}

In this paper, we study the following research questions:

\newcommand\rqone{What is the provenance of the nearest neighbor diff?}

\newcommand\rqtwo{What is the performance difference between inter- and intra-repository nearest-neighbor-based commit message generation?}

\begin{itemize}
    \item RQ1: \rqone \ Previous research on using the nearest neighbor algorithm for commit message generation has completely overlooked the question of the repositories containing the nearest neighbor. We fill this gap and provide original and unique results on the nearest diff provenance.
    \item RQ2: \rqtwo \ Since the provenance of nearest neighbor diffs is not studied, it is still unknown how the results are affected if the learning is focused on each of the origins. We study the performance of nearest-neighbor-based commit message generation techniques when inter-repository learning, intra-repository learning, or both are used. 
\end{itemize}

\subsection{Dataset}
\label{sec:dataset}
The dataset used in this paper comes from previous work \cite{jiang2017automatically,liu2018neural}. The original data has been collected from top $1k$ Github \footnote{https://github.com/} repositories and contains over $2M$ commits (\textit{original dataset}). After applying several filters, Jiang et al. obtain a set of $32k$ commits that could be used by NMT1 algorithm \cite{jiang2017automatically}.

Liu et al. made a cleaned version of this dataset by removing the \textit{noisy messages} \cite{liu2018neural} (\textit{cleaned dataset}). The noisy messages are categorized into two categories: 
\begin{enumerate*}
    \item The messages generated by development tools (called \textit{bot messages}).
    \item The messages containing little and redundant information (called \textit{trivial messages}). 
\end{enumerate*}
In total, about 16\% of the messages are considered as noisy. As a result, the cleaned dataset includes $22112$, $2511$, and $2521$ commits in the training, validation, and testing subsets, respectively. Note that since NNGen, in contrast with NMT1, does not build a model or tune any hyperparameters, the validation subset is not used by it.

\begin{figure}[!t]
\includegraphics[width=9.2cm]{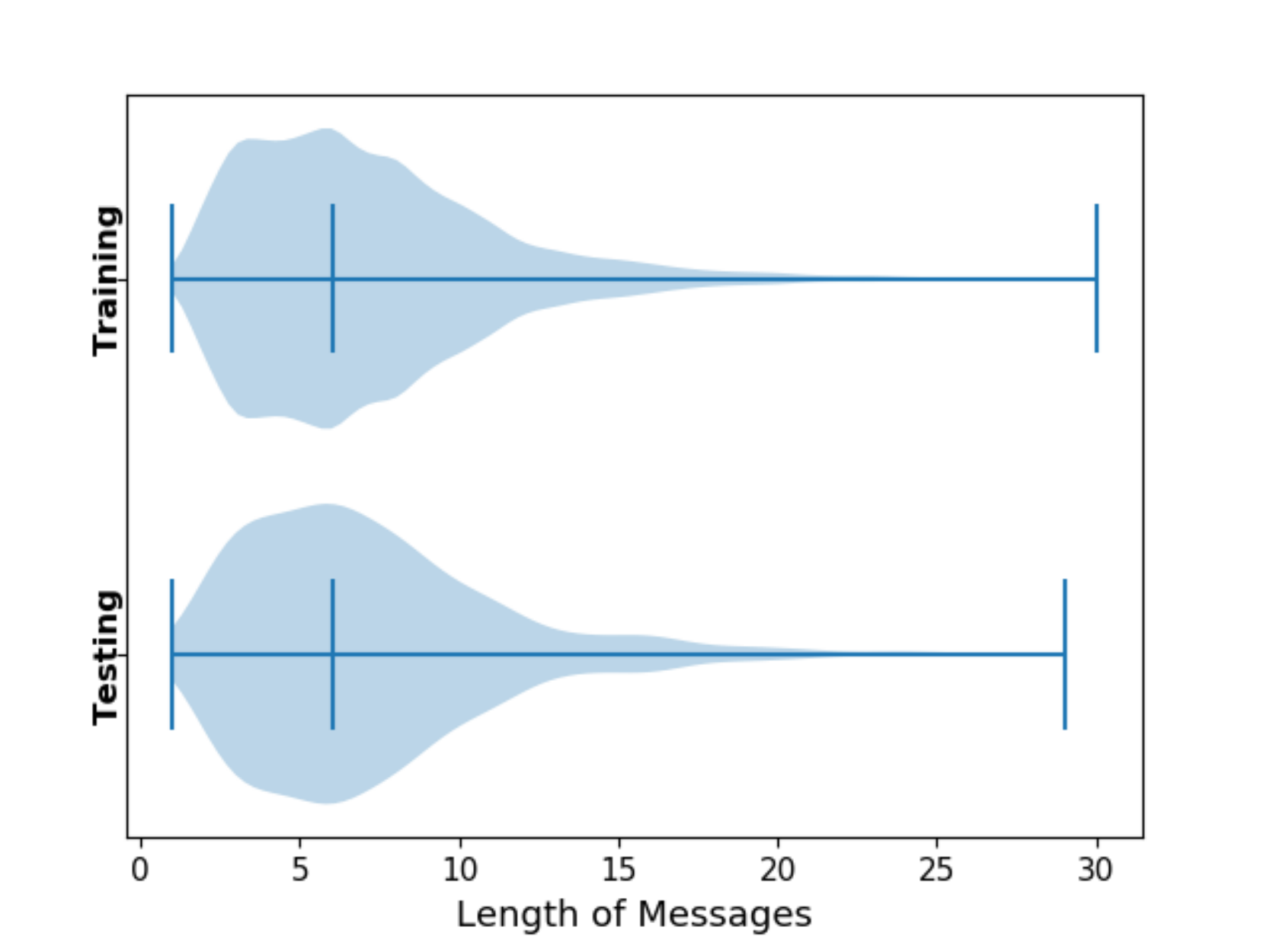}
\caption{The distribution of the number of words of commit messages for each subset of the filtered dataset.}
\label{fig-msg-length}
\end{figure}

\begin{figure}[!t]
\includegraphics[width=9.2cm]{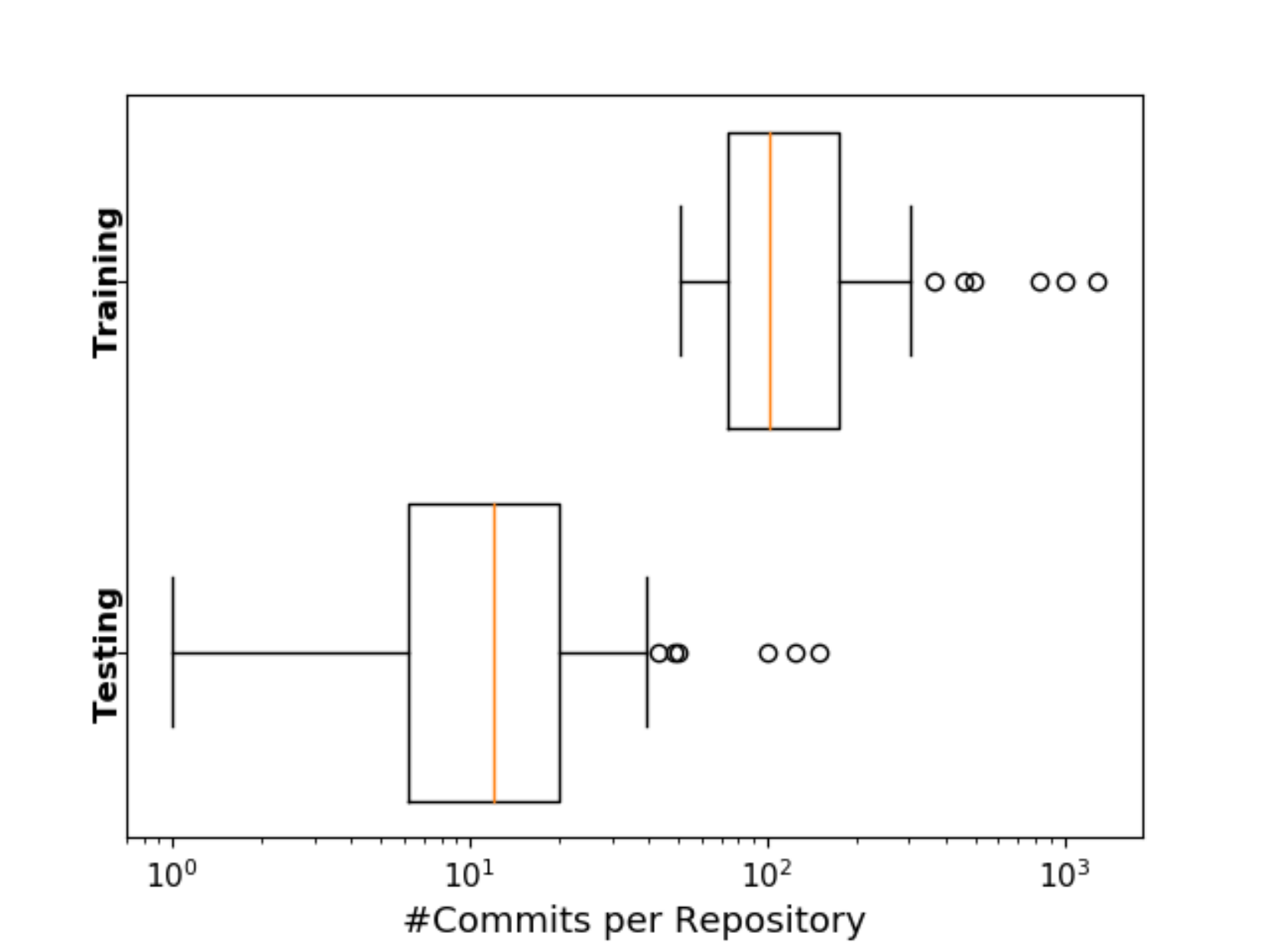}

\caption{The distribution of the number of commits per repository for each subset of the filtered dataset.}
\label{fig-repo-commits}
\end{figure}

In this work, we define the provenance of a nearest neighbor diff as the repository containing it. Therefore, we aim at finding the containing repository for each commit in the training and testing datasets. This data is not included in the dataset shared by previous research papers \cite{jiang2017automatically,liu2018neural}. Hence, we enrich the cleaned dataset by collecting this information as follows. We search for each cleaned commit message in the set of $2M$ extracted messages in the non-preprocessed original dataset. Therefore, we find the non-preprocessed commit that corresponds to each commit in the cleaned datset. In the non-preprocessed dataset (but not in the cleaned dataset), each message is mapped to a repository\_id and a commit\_id. As a result, we are able to map each commit from the cleaned dataset to a Github repository address. However, a cleaned commit message might be identical to more than one messages of the non-preprocessed dataset. In such cases we take the first one, and consequently, some of the extracted mappings might be incorrect.

In order to make sure that the generated mappings between commits and  corresponding repositories are accurate enough, we randomly select and investigate 30 mappings. This manual investigation shows that 29 commits are mapped to the actual Github repository they belong to. Only in one case out of thirty, where the message is ``Fix typo'' (a generic message), the mapping is not accurate. Hence, we conclude that the extracted mappings are reliable enough for further studies. We should also note that for $2\%$ of the messages the corresponding repository could not be found using our method. This happens because some of the messages are slightly changed in the cleaned dataset compared to the original version. The dataset augmented with this mapping information is made publicly available for future research: \url{https://github.com/khaes-kth/simple-nngen}.

To have a dataset on which intra-repository learning can be meaningfully evaluated, every repository in the dataset should contain more than a minimum number of commits. In this manner, we create a \textit{filtered dataset} by removing the data from repositories with 50 or less commits in the training dataset. This dataset is used to answer RQ2 in \autoref{sec:protocol-rq2}. The filtered dataset contains $14738$ training commits and $1665$ testing commits. The \autoref{fig-msg-length} shows the distribution of commit messages length in number of words for this dataset. The median length for a message is six words in both of the subsets. An example for a message with six words is ``Remove fonts from sysui package .'' which is collected from Android \textit{platform\_frameworks\_base}\footnote{\url{https://github.com/aosp-mirror/platform_frameworks_base}} repository.

\subsection{Protocol for RQ1}

``\rqone''

In answer to RQ1, we reproduce the setup of \cite{liu2018neural}. For each test diff, we find the five training diffs with highest cosine-similarity to the test diff and select the one with highest BLEU\_4 score as the nearest neighbor. In addition to this setup, we trace and measure the repository of nearest neighbor diffs. 

Afterwards, we measure $ORG\_R$, the ratio of the nearest diffs that NNGen selects from a specific origin $R$. In this regard, $ORG\_R_{samerepo}$ and $ORG\_R_{otherrepo}$ represent the ratio of nearest diffs that NNGen selects from the same and different repositories, respectively.

Moreover, in order to assess the quality of messages selected from the same/different repository, we compute the BLEU\_4 score between each generated message and the corresponding reference message. We then calculate $MBLEU$, the mean BLEU\_4 score for the generated messages selected per repository origin, that is we compute $MBLEU_{samerepo}$ and $MBLEU_{otherrepo}$.

\subsection{Protocol for RQ2}
\label{sec:protocol-rq2}

``\rqtwo''

To answer RQ2, we design two variations of the NNGen method: Simple-NNGen and EXC-NNGen and measure their performance.
\textbf{Simple-NNGen} searches for the nearest training diff only in the repository of the given test diff.
\textbf{EXC-NNGen} searches for nearest diffs only in the repositories other than the repository of the given diff.
Simple-NNGen and EXC-NNGen represent typical intra- and inter-repository nearest-neighbor-based commit message generation techniques, respectively.

As for RQ2, we compute the BLUE\_4 scores between the human-written messages and the output of Simple-NNGen and EXC-NNGen, and NNGen. A higher BLUE\_4 score between a set of generated messages and human-written messages indicates a higher quality for the generated messages. Therefore, if Simple-NNGen has a higher BLEU\_4 score, it would mean that intra-repository nearest-neighbor-based commit message generation outperforms the inter-repository version of this technique.

\section{Experimental Results}

\subsection{\rqone}

As shown in the ``ORG\_R'' column of \autoref{tab:scores_per_origin}, we find that for $60\%$ ($1520/2521$) of the test diffs, the most similar diff selected by NNGen comes from the same repository. In other words, for $60\%$ of the cases, NNGen uses no information from repositories other than the repository of test commit. This suggests that NNGen does not take advantage of cross-project learning effectively. Note that ``unknown repository'' origin in this table represents commits for which the repository could not be retrieved using the method explained in \autoref{sec:dataset}.

The MBLEU scores over 2521 generated messages that are selected from the same and different repositories are $13.13$ and $3.29$, respectively (the higher, the better). Therefore, this means that for the test diffs for which NNGen picks the closest diff from the same repository, the generated message is significantly better because it is closer to the ground-truth message.

\begin{framed}\noindent
    \textbf{Answer to RQ1: \rqone} \\
    For $60\%$ of the test commits, the message generated by NNGen is selected from the same repository and no information from other repositories is used. The mean BLEU\_4 score between these messages and reference messages is $13.13$ while the same measure for the messages selected from other repositories is $3.29$, showing that the quality of the generated message is much higher when only intra-repository learning is used.
    
    These results indicate that a nearest-neighbor-based algorithm might be as effective as NNGen when it only uses the information from the repository of the test commit. The Simple-NNGen technique that will be assessed in the next RQ is designed based on this observation.
    
\end{framed}

\subsection{\rqtwo}

\begin{table}[t]
\centering
\caption{RQ1: Number of nearest neighbors and MBLEU score per origin.}
\label{tab:scores_per_origin}
\begin{tabular}{c|c|c}
\hline
	\textbf{Origin} & \textbf{ORG\_R} & \textbf{MBLEU} \\ \hline \hline
	full test dataset & 2521 & 9.30 \\ \hline

	same repository & 1520/2521 (60\%) & 13.13 \\ \hline
	other repository & 949/2521 (38\%) & 3.29 \\ \hline
	unknown repository & 52/2521 (2\%) & 7.03 \\ \hline
\end{tabular}
\end{table}

\begin{table}
\centering
\begin{threeparttable}
\centering
\caption{RQ2: BLEU\_4 Scores on 1665 commit generation tasks. The higher, the better.}
\label{tab:bleu_scores}
\begin{tabular}{c|c|c|c|c|c}
\hline
	\textbf{Method} & \textbf{BLEU\_4} & \textbf{$p_1$} & \textbf{$p_2$} & \textbf{$p_3$} & \textbf{$p_4$}\\ \hline \hline

	NNGen & 17.06 & 28.5 & 17.7 & 14.1 & 12.5 \\ \hline
	EXC-NNGEN & 2.68 & 10.8 & 2.9 & 1.9 & 1.7 \\ \hline
	SIMPLE-NNGEN & 17.64 & 28.8 & 18.1 & 14.5 & 12.8 \\ \hline
\end{tabular}
\begin{tablenotes}
  \small
  \item $p_n$ ($n=1,2,3,4$) refers to the modified n-gram precision.
\end{tablenotes}
\end{threeparttable}
\end{table}

\autoref{tab:bleu_scores} shows the results of our second experiment conducted to answer RQ2. This table includes the BLEU\_4 scores for messages generated by NNGen, Simple-NNGen and EXC-NNGen. Bear in mind that in order to make sure intra-repository learning is evaluated on a large enough dataset, we only considered messages generated for the test diffs included in the filtered dataset. As explained in \autoref{sec:dataset}, the data from repositories with 50 or less training commits are removed in this dataset.

The remarkable difference between the performance of EXC-NNGen and Simple-NNGen indicates that the intra-repository nearest-neighbor-based technique is superior to the inter-repository version of this algorithm. The results also suggest that Simple-NNGen outperforms NNGen. Furthermore, since \cite{liu2019atom} shows that NNGen outperforms all existing NMT-based methods, one might conclude that Simple-NNGen also outperforms NMT1 and more complex versions of neural-machine-translation based techniques. Consequently, Simple-NNgen could be used as a simple, fast, and effective baseline algorithm for future studies.

Note that NNGen searches for the nearest neighbor among all the $22112$ commits included in the cleaned training dataset, while Simple-NNGen only considers the commits from the same repository. The distribution of the number of commits per repository is shown in \autoref{fig-repo-commits}. The median number of training commits per repository is $102.5$. This indicates that Simple-NNGen requires notably shorter learning process compared to NNGen.

Despite all the promising results presented in this paper, there is a threat to the validity of our experiment. The threat is that since Simple-NNGen relies on the commit history of the repository of the given test commit, it might fail if the repository does not contain a large enough commit history. The median number of training commits per repository is $102.5$ in our experiment which means we may not be able to generalize our findings to newly created repositories with very small histories. Moreover, we know that the included repositories are from the top $1k$ projects on Github which makes them likely to use well-defined structures for their commit messages. Therefore, Simple-NNGen might also not be able to perform well for repositories without established commit message structures. More empirical study is needed to address this concern in the future.

\begin{framed}\noindent
    \textbf{Answer to RQ2: \rqtwo} \\
    Simple-NNGen is a variation of NNGen with the difference that it searches for the most similar diff only in the commit database from the same repository. Simple-NNGen outperforms NNGen and EXC-NNGen in terms of the BLEU\_4 score. To put it more clearly, it shows that an intra-repository method outperforms inter-repository methods, showing the difficulty of cross-repository learning with nearest neighbours for commit message generation.
\end{framed}

\section{Related Work}

In addition to NMT1 \cite{jiang2017automatically} and NNGen \cite{liu2018neural}, other commit message generation techniques have been introduced in recent years. In this section, we briefly review these techniques.

\subsection{Rule-based Methods}
There are studies using predefined rules or templates for commit message generation. In 2010, Buse and Weimer introduced DeltaDoc\cite{buse2010automatically}. DeltaDoc produces summaries that are longer than normal commit messages but shorter than raw diffs. This approach consists of three steps. In the first step, for each statement, the conditions under which that statement can be executed are extracted. In the second step, for each old statement like Z that would be executed under a condition like X and is changed to a new statement like Y, a document of the form ``When calling A(), If X, Do Y Instead of Z'' is generated. Here, A() is the function containing Y. In the final step, the generated document is summarized using several heuristics.

The \textit{ChangeScribe} tool \cite{linares2015changescribe} and the technique by Shen et al. \cite{shen2016automatic} both use method stereotypes distribution \cite{dragan2006reverse} in the commit to determine the commit type. The determined commit type is supposed to answer the question ``\textit{why} the source code is changed in this commit?'' Regarding this why question, the method presented by Shen et al. \cite{shen2016automatic} also identifies the maintenance type of the change (ex., corrective, perfective) and reports it in the generated commit message. Both of these works also detect and report the parts of the code that are changed in order to answer the question ``\textit{what} is changed in this commit?''

\subsection{Neural-machine-translation-based Methods}
Another trend in the commit message generation research field is using neural machine translation for message generation. In this regard, NMT1 \cite{jiang2017automatically} can be seen as one of the first works that report large-scale results. Earlier works on that idea include those of Vishalakshi and Krishnapriya \cite{Vishalakshi2016AutomaticGO} and Loyola et al. \cite{loyola2017neural, loyola2018content}.

There are also more advanced NMT-based commit message generation methods. PtrGNCMsg \cite{liu2019generating} is a new approach that addresses the out-of-vocabulary (OOV) problem by using pointer-generator network. In this manner, before generating each of the commit message words, PtrGNCMsg uses probabilistic techniques to decide whether the word should be selected from the training vocabulary or copied from the given test commit diff.

CODISUM\cite{xu2019commit} and ATOM\cite{liu2019atom} are commit message generation methods that also take the code change structural information into account to generate the messages. In this regard, CODISUM replaces class/method/variable names with placeholders to extract the structural information of source code changes.

On the other hand, ATOM sees the code change diffs as changes in the abstract syntax trees (AST) of programs instead of plain text. Besides proposing an NMT-based technique for generating a commit message, ATOM\cite{liu2019atom} presents a retrieval technique which finds the most relevant training diff to a given test diff in terms of the tf-idf score \cite{aizawa2003information}. Finally, it ranks the generated and retrieved messages and returns the better one as the output. This hybrid algorithm outperforms both NNGen and NMT-based algorithms in terms of BLEU\_4 score.

\subsection{Retrieval-based Methods}
NNGen \cite{liu2018neural} can be seen as a retrieval-based approach to retrieve the most similar training diff to the test diff.
An other retrieval-based method is the one introduced in \cite{huang2017mining}. This work employs a token-based method \cite{wettel2005archeology} to calculate the syntax similarity between the test diff and training diffs. It also computes the semantic similarity between two documents w.r.t. the vectorial angle of their semantic vectors that are built using latent semantic analysis. Finally, the calculated syntax and semantic similarities are used to find the most similar training commit diff. Since Huang et al.'s approach \cite{huang2017mining} calculates the syntax similarity, it only works for the diffs that merely consist of source code changes. However, NNGen, NMT1, and Simple-NNGen work for all kinds of diffs which makes them more general solutions to the commit message generation problem.

\section{Conclusion}
In this paper, we investigate the origin of nearest diffs selected by NNGen. By doing so, we try to understand NNGen strengths and drawbacks. We find that for $60\%$ of test diffs, the nearest training diff selected by NNGen comes from the same repository as the test diff repository. Moreover, the mean BLEU\_4 score between the generated messages that NNGen selects from the same repository is 13.13, while this measure for other messages generated by NNGen is 3.29, meaning that NNGen performs significantly better when it ignores training commits from other projects. Based on these observations we conclude that NNGen does not take advantage of cross-project learning. We then introduce Simple-NNGen which is a variation of NNGen that only searches among the diffs from the same repository to find the nearest diff. Our evaluations indicate that Simple-NNGen outperforms NNGen, which shows it can be used as a new baseline in future research.
In the future, we plan to carry out a comprehensive empirical study on the relation between commit history properties and the performance of Simple-NNGen and other commit message generation methods.

\begin{acks}
This work was partially supported by the Wallenberg Artificial Intelligence, Autonomous Systems and Software Program (WASP) funded by Knut and Alice Wallenberg Foundation
and the Swedish Foundation for Strategic Research (SSF).
\end{acks}

\bibliographystyle{ACM-Reference-Format}
\bibliography{references}

\end{document}